# Spinning Pupil Aberration Measurement for anisoplanatic deconvolution

**Daniele Ancora,**[1] **Tommaso Furieri,**[2,3] **Stefano Bonora,**[2,*] **Andrea Bassi**[1]

[1]*Politecnico di Milano, Department of Physics, piazza Leonardo da Vinci 32, 20133 Milan, Italy*
[2]*National Council of Research of Italy, Institute of Photonics and Nanotechnology, via Trasea 7, 35131, Padova, Italy*
[3]*University of Padova, Department of Information Engineering, Via Gradenigo 6, 35131, Padova, Italy*

*\*Corresponding author: stefano.bonora@pd.ifn.cnr.it*

**The aberrations in an optical microscope are commonly measured and corrected at one location in the field of view, within the so-called isoplanatic patch. Full-field correction is desirable for high-resolution imaging of large specimens. Here we present a novel wavefront detector, based on pupil sampling with sub-apertures, which measures the aberrated wavefront phase at each position of the specimen. Based on this measurement, we propose a region-wise deconvolution that provides an anisoplanatic reconstruction of the sample image. Our results indicate that the measurement and correction of the aberrations can be performed in a wide-field fluorescence microscope over its entire field of view.**

The measurement of wavefront aberration is of crucial interest in optical microscopy, which aims to image samples at the highest possible resolution. The wavefront measurement can be carried out with several methods, for example with the use of wavefront sensors (e.g. Shack-Hartman wavefront sensors), by the acquisition of the system's Point Spread Function (PSF) at different depths [1] or indirectly by using a pupil segmentation method with the use of DMD or SML [2][3]. In fluorescence microscopy, the PSF can be directly measured by imaging a point like source, such as fluorescent nano-beads, smaller than the diffraction limit of the optical system. When a good knowledge of the PSF is available, deconvolution is effectively used to provide a reconstruction of the object at high resolution [4]. However, the isoplanatic patch, i.e. the area where the aberration can be considered constant, is often smaller than the field of view. This happens when the aberrations are developed on different planes in the volume between the sample and the objective [5]. In this case, the PSF is not isoplanatic and changes in different imaged regions. Wavefront sensors measure the aberrations with high accuracy only in one point. Consequently, the reconstruction of the wavefront aberration corresponding to different regions of the sample would require the presence of many bright point sources in the field of view. This technique is successfully used in astronomy [6] and in some cases in ophthalmic imaging [7]. Similarly, the direct measurement of the PSF in different locations, when possible, is affected by noise and requires the localization of the fluorescence emitters that are sparse in different regions of the sample. Therefore, the presence of an anisoplanatic PSF makes the use of deconvolution algorithms critical [8,9] since an incorrect knowledge of the PSF creates artifacts in the reconstruction.

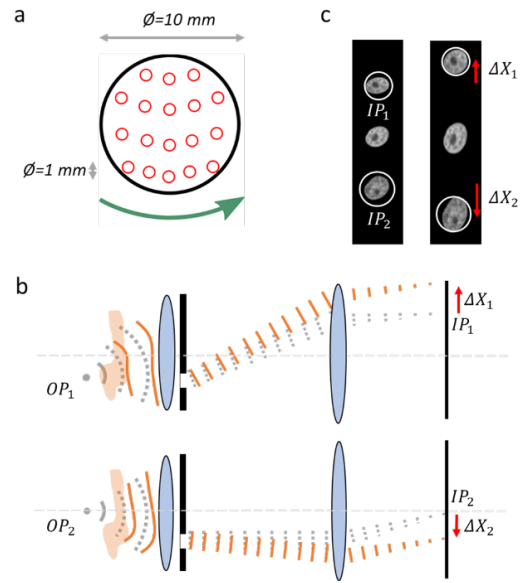

**Fig. 1**. Pupil of the optical system (black circle) and position of the sub apertures (red circles) in the pupil (a), (see Visualization 1). Schematic diagram of the optical imaging system with an example of image shift relative to two fields on the object ($OP_1$ and $OP_2$) (b). Example of the image distortion and shift due to the presence of a wavefront aberration (c).

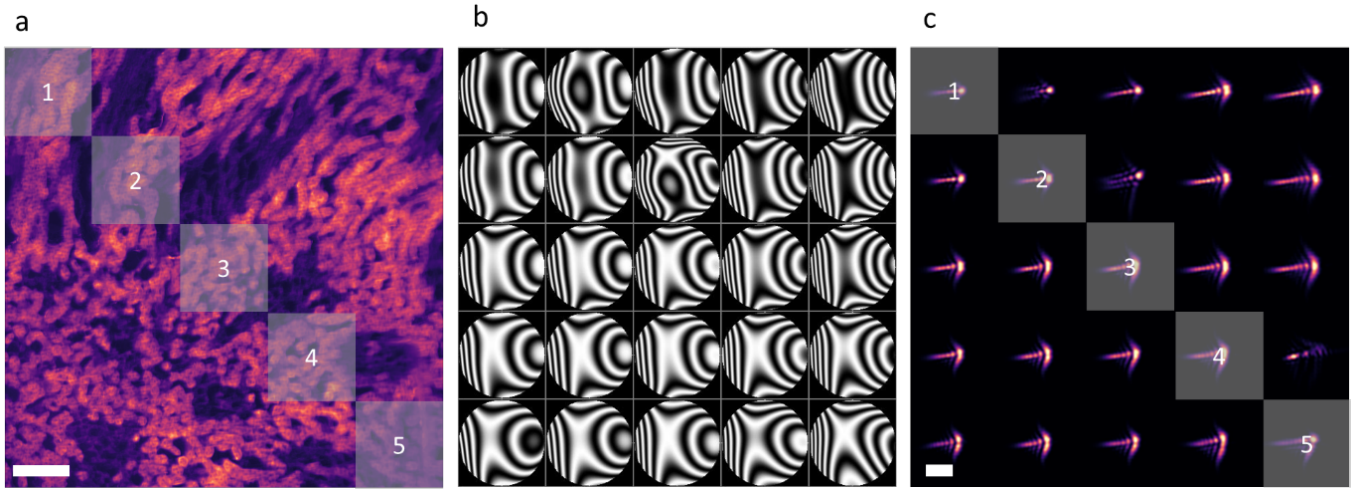

**Fig.2**. Acquired image of a mouse kidney slide, aberrated by the presence of a glass wedge (a). Reconstructed wavefronts in 5x5 subregions (b) and corresponding PSFs (c). In the actual reconstructions, we mapped the wavefront in a 9x9 grid, whereas here we show a 5x5 for a visual purpose only. Scale bar is 200μm in (a) and 10 μm in (c).

In this letter we first introduce a new method for wavefront measurement able to reconstruct the wavefront aberration and the PSF corresponding to all points of the field of view. Then, we present a deconvolution method that, starting from the measured PSFs, reconstructs the samples anisoplanatically in its different regions. Using this method, we show high resolution, artifacts-free reconstruction of a sample slide in a widefield fluorescence microscope.

Our wavefront detection method is based on a Spinning sub-Pupil Aberration Measurement (SPAM). The measurement starts with a motorized device that moves a sub-aperture across the microscope's pupil. For each position of the sub-apertures (Fig. 1a), an image of the sample is acquired with the camera of the widefield microscope (Visualization 1). By measuring the relative shift of each acquired image, the wavefront gradient in each pupil location is obtained. Figure 1b shows this process for an object point $OP_1$. When imaged through a certain sub-aperture, its image $IP_1$ is shifted by a quantity $\Delta x_1$ from the non-aberrated position. Since we do not have access to the non aberrated image, we choose as the reference one, the image acquired with a sub aperture placed in the centre of the pupil. The displacement $\Delta x_1$ is thus proportional to the wavefront gradient in the corresponding pupil position. The shift is then measured using the Sum of Squared Difference Algorithm [10]. The wavefront reconstruction is equivalent to the Shack-Hartmann procedure (see Supplemental document). It is worth noting that for the same sub-aperture position, different fields on the object ($OP_2$) can suffer from different wavefront aberrations, potentially leading to a different image shift $\Delta x_2$. Thanks to this principle the SPAM can measure the wavefront in each position of the field of view. Our wavefront measurement system is composed of a spinning wheel that, rotating, scans the pupil with an aperture of 1 mm across a pupil of 10 mm diameter. This configuration was chosen to obtain a good trade-off between wavefront sampling and device compactness and allows the measurement of aberrations up to the 4$^{th}$ order of Zernike polynomials (OSA/ANSI j = 14). The scan happens on 4 radial distances, with the angular positions of the iris being designed to avoid overlaps (Fig. 1a). The total scanned points on the wavefront are 18 and, consequently, 18 images are acquired. The camera automatically adjusts the exposure time proportionally to the ratio R of the area of the sub-apertures and of the whole pupil (R = 100). The spinning wheel is also equipped with an aperture, as large as the pupil of the microscope, which is used for image acquisition. For a typical fluorescent sample slide, the exposure time with the fully open pupil is 10ms and it is 1s for the sub-apertures. The total time required for the wavefront measurement is 18s. It is worth noting that the sub-apertures could be realized with a bigger diameter for increased light collection during the scanning process but at the expense of wavefront measurement accuracy. The images are acquired with a 10X microscope (Mitutoyo, Plan Apo, LWD, NA=0.28) objective in a custom made widefield upright microscope, equipped with a 200 mm tube lens (Nikon) and a sCMOS camera (Hamamatsu Flash 4.0, 2048 × 2048 pixels).

Figure 2 shows the image of a microscope slide (FluoCells, Thermofisher, Prepared Slide #3, 16 μm mouse kidney section labelled with Alexa Fluor 488). An aberrating phase plate, made with two 170 μm cover-glasses bonded with UV glue, is placed in front of the sample, inclined by 45° to induce a large aberration. A 480nm Light-Emitting Diode illuminates the sample and the light is detected with a 520±15nm filter placed in between the SPAM module and the tube lens. Thus, the SPAM reconstructs the region-dependent wavefront phases (Fig. 2b) in $N \times N$ regions (here $N=9$). The corresponding PSFs are calculated as the squared value of the 2D Fourier Transform of the wavefront phases (Fig. 2c). We observe that the aberrations contain mainly astigmatism and coma. Since we have access to the PSF in the different locations of the sample, we developed a region-wise deconvolution, inspired by an approach proposed in astronomy [11]. We implement the deconvolution using a Richardson-Lucy (RL) iterative method [12]. A simple deconvolution for each of the $N \times N$ regions of the image would produce artifacts at the boundary of each region. These artifacts will typically have a size comparable to the one of the PSF. Moreover, each tile abruptly changes the PSF, possibly leading to a discontinuous reconstruction. Assuming that the PSF changes over the image with continuity, we use the following approach to avoid the formation of stitching artifacts. We divide the camera image, into overlapping tiles, defining a window size of $W$ pixels (here $W$ = 410 px and a step (or stride) of $s$ = 82 px. The stride was set to be comparable with the size of the PSF images and to divide the image with an integer number. This choice produces a map of $M \times M$ tiles ($M$ = 21). Adopting a notation similar to that used in the field of convolutional neural networks, we rearrange the image-data in channels (each tile is a channel in this notation), forming a matrix with dimensions $C \times W \times W$. The tiled view of the original image, $T$, that has a total of $C = M^2 = 441$ tiles (Fig. 3a). To match the problem size, we expand the PSF map from $N \times N$ to the reach the same tiled dimension of the image $M \times M$, by oversampling it via bicubic interpolation. The oversampled PSF are also rearranged, forming a kernel of $C \times W' \times W'$, where $W'$ indicates the dimension of each PSF. In our case study, $W'$ = 93 px. In this formalism, each channel of the image is blurred by the corresponding channel of the PSF map, according to:

$$T(c,x,y) = D(c,x',y') *_{(x',y')} \text{PSF}(c,x',y')$$

where, $D$ is the deconvolved tiled image and the convolution operator acts only on the last two dimensions. Once we are settled with this expanded view of the image $T$ and the PSF, we can formulate the iterative RL deconvolution as:

$$D^{i+1} = D^i \left( \frac{T}{D^i *_{(x',y')} \text{PSF}} *_{(x',y')} \widetilde{\text{PSF}} \right)$$

where $D^{i+1}$ is the deconvolved dataset after $i+1$ steps and $\widetilde{\text{PSF}}(c,x',y') = \text{PSF}(c,-x',-y')$. This formulation closely resembles the action of two convolutional layers (with a division in between) applied depth-wise along the channel direction. For this reason, we implemented the code by using convolution functions from the GPU accelerated PyTorch package, commonly used in machine learning [13]. After the RL deconvolution, we are left with a dataset $D$ (Fig. 3) and, to reconstruct the final image, we un-tile this dataset. From each tile, we suppress an external frame (corresponding to 32px). We perform this operation to remove the region where the RL algorithm produces artifacts. Then, we place each tile back into its corresponding location, averaging the overlapping regions. This is done by counting the number of times that each pixel was included in the tiled compositions. Repeating this procedure through the whole image plane, leaves us with the recomposed deconvolved image (Fig. 3c). We compare the results of a standard, isoplanatic (ISO), deconvolution and the proposed region-wise, anisoplanatic (ANI) deconvolution in Fig. 4. For this comparison, we have set the same number of iterations (200) for any reconstruction presented in this letter. In the case of ISO, we use the PSF retrieved in the centre of the field of view, to deconvolve the whole image. Consequently, the results of ANI and ISO are perfectly similar in the image centre (data not shown). Both deconvolutions can restore an image with improved contrast than the acquired one: comparing a detail of the original image (Fig. 4c) with the deconvolutions (Fig. 4e and Fig. 4f), a substantial deblur is observable. However, observing the peripherical regions of the field of view, we note that ISO produces artifacts on the image, as typically happens when performing deconvolution with an inaccurate PSF. This is noticeable by comparing the ISO reconstruction (Fig. 4e) with the image of the sample acquired without the aberrating glass (Fig. 4d). We observe that ISO reconstructed image, is systematically affected by artifacts: as an example, Fig. 4e show that several structures appear in the proximity of the bright locations, which are not present in the biological sample (Fig. 4d). The arrows indicate some locations in which this type of artifacts is particularly marked. Conversely ANI reconstruction provides an improved contrast and deblurring without generating structures that are not present in the original specimen. Our deconvolution is inspired by modern concepts in parallel data processing and permits a straightforward gain in performance offered by GPUs. Finally, the combination of anisoplanatic deconvolution and adaptive optics [14] could be the key for achieving perfect imaging over extended areas of observation.

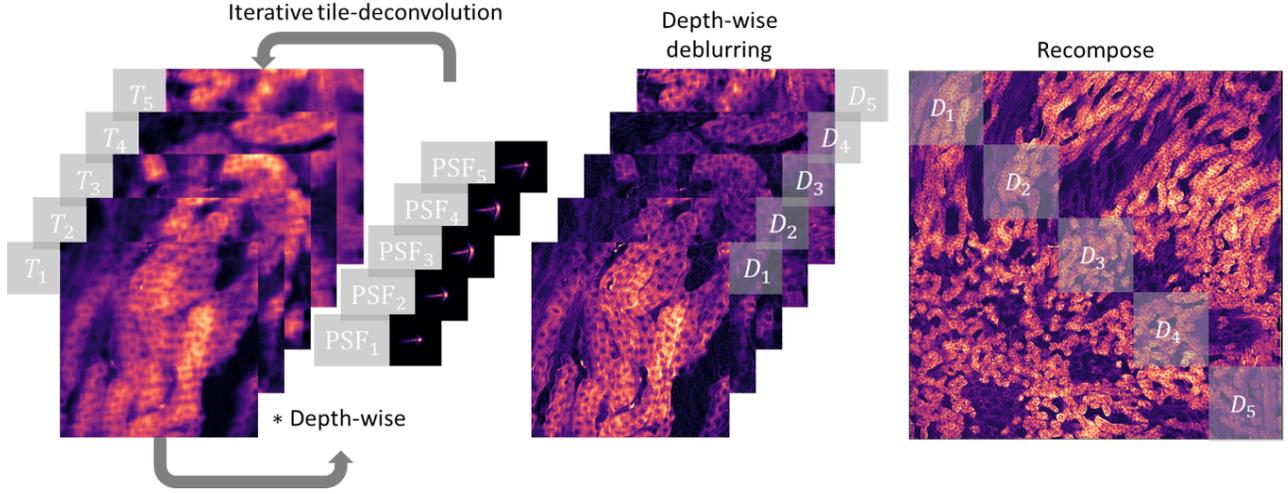

**Fig. 3**. Schematics of the ANI deconvolution problem. A tiled view T of the image t (on the left side) is deconvolved via iterative RL steps with its corresponding PSF channel. This produces a deblurred tiled image D (central panel), which is, then, recomposed to form the final deconvolved image on the right side.

In conclusion, we have proposed a novel method to measure the wavefront aberrations by sampling the pupil of an optical imaging system, particularly of a widefield microscope. The method allows reconstructing the wavefront for the different points of the field of view. The SPAM module can be inserted in the detection path of a widefield microscope and does not require the use of external wavefront sensors. It provides PSF reconstruction, by imaging the sample under the microscope, without the need of using beads or calibration targets. In combination with the wavefront and PSF measurement, we have developed a multi-region, deconvolution method, that takes into consideration the non-isoplanatic PSF and carries out a region-dependent iterative deblurring. The aberration measurement, along with the proposed deconvolution has shown to be effective in reconstructing fluorescence imaging samples at high resolution, avoiding artifacts otherwise present in a conventional isoplanatic deconvolution.

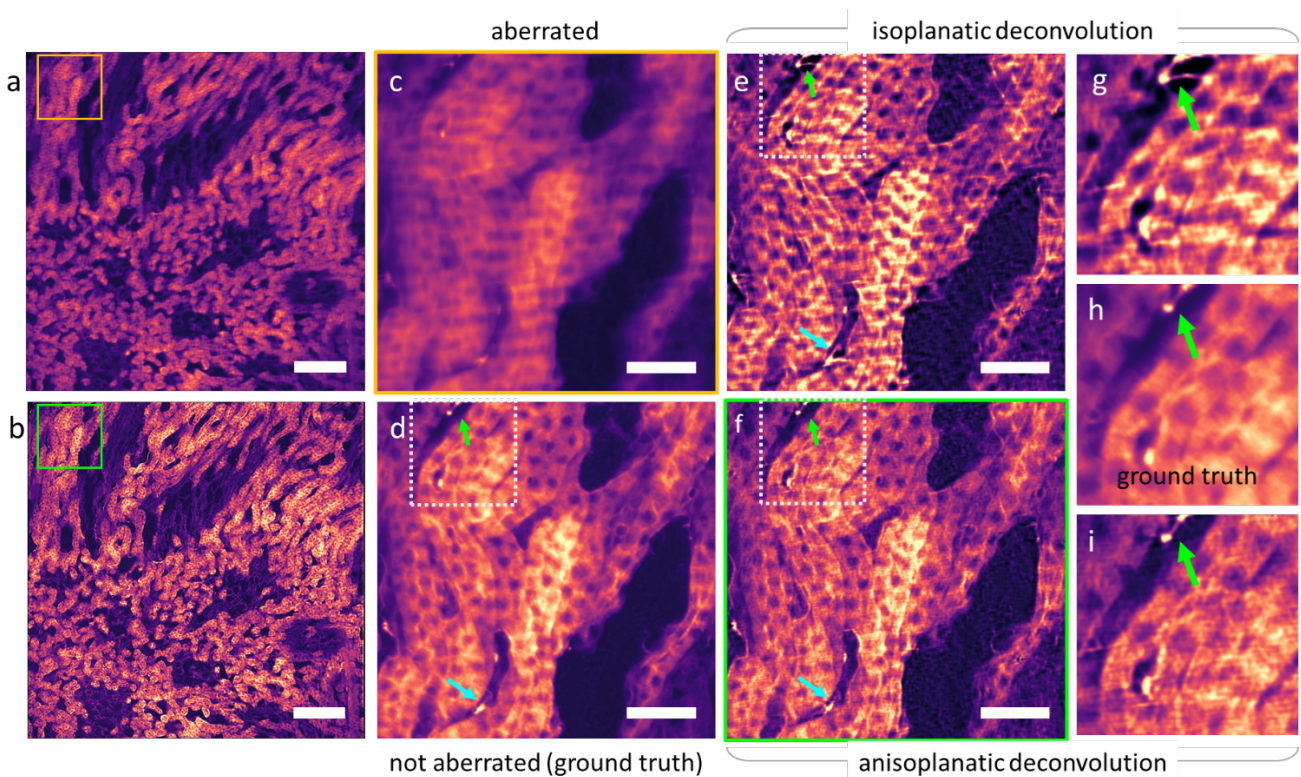

**Fig. 4.** Acquired (a) and ANI deconvolved (b) image of a mouse kidney microscope slide. Magnified detail (corresponding to the squared inset in panels a,b) of the image acquired in the presence of the glass aberration (c), ground truth acquired without the aberrating phase plate (d), deconvolved with ISO (e), deconvolved with ANI (f). Further magnified detail (corresponding to the squared inset in panels e,f) of the reconstruction with ISO (g), the ground truth (h), and ANI (i). Some artifacts created by ISO deconvolution are indicated with the green and light blue arrows. Scalebar is 200μm in (a,b) and 50 μm in (c-f).

**Acknowledgments**

CNR-IFN was supported by Office of Naval Research global (ONRG) and by Air Force Research Laboratory (AFRL) with grant GRANT12789919. Politecnico di Milano has received funding from LASERLAB-EUROPE (grant agreement no. 871124, European Union's Horizon 2020 research and innovation programme) and from H2020 Marie Skłodowska-Curie Actions (HI-PHRET project, 799230).

**Disclosures.** The authors declare no conflicts of interest.

**Supplemental document**. See Supplement 1 for supporting content